\documentclass{svjour2}
\usepackage{graphicx}
\usepackage{latexsym}
\usepackage{amssymb}
\usepackage{wasysym}
\RequirePackage{color}

 \def\rmd{{\rm d}}

\def\beq{\begin{equation}}
\def\eeq{\end{equation}}
\def\pmb#1{\setbox0=\hbox{$#1$}%
  \kern-.025em\copy0\kern-\wd0
  \kern.05em\copy0\kern-\wd0
  \kern-.025em\raise.0433em\box0}

\def\dal{\mathop{\rlap{\hbox{$\sqcap$}}\sqcup}\nolimits}   

 \def\cu{\cos u}
 \def\su{\sin u}

 \def\nk{n_{\rm b}}

\def\dert#1#2{\frac{{{d}}{#1}}{{{d}}{#2}}}
\def\kap{\hat{\mathbf S}}

\def\co{\cos\omega}
\def\so{\sin\omega}
\def\coo{\cos 2\omega}
\def\soo{\sin 2\omega}
\def\cO{\cos\mathit{\Omega}}
\def\sO{\sin\mathit{\Omega}}
\def\cI{\cos I}
\def\sI{\sin I}

\def\ton#1{\left(#1\right)}
\def\qua#1{\left[#1\right]}

\journalname{General Relativity and Gravitation}

\begin{document}

\title{Orbital effects due to gravitational induction }

\author{Donato Bini\and
Lorenzo Iorio\and
Domenico Giordano
}

\institute{
Donato Bini
\at
Istituto per le Applicazioni del Calcolo ``M. Picone,'' CNR, I--00161 Rome, Italy\\
\email{binid@icra.it}
\and
Lorenzo Iorio
\at
Ministero dell'Istruzione, dell'Universit$\grave{\textrm{a}}$ e della Ricerca\\
Fellow of the Royal Astronomical Society (F.R.A.S.)\\
Viale Unit$\grave{\textrm{a}}$ di Italia 68, 70125, Bari (BA), Italy\\
\email{lorenzo.iorio@libero.it}
\and
Domenico Giordano
\at
Aerothermodynamics Section ESA - ESTEC
Keplerlaan 1, 2201 AZ  Noordwijk, The Netherlands\\
\email{Domenico.Giordano@esa.int}
}

\date{Received: date / Accepted: date / Version: \today}

\maketitle
\begin{abstract}
We study the motion of test particles in the metric of a localized and slowly rotating astronomical source, within the framework of  linear gravitoelectromagnetism, grounded on a Post-Minkowskian approximation of  general relativity. Special attention is paid to gravitational inductive effects due to time-varying gravitomagnetic fields. We show that, within the limits of the approximation mentioned above,  there are cumulative effects on the orbit of the particles either for planetary sources  or for binary systems. They turn out to be negligible.
\end{abstract}

\section{Introduction}

Einstein's gravitational theory, \lq\lq general relativity," is largely accepted as the best description of the gravitation interaction  available today.
Apart from classical test dated back to the beginning of the last century, the main expectation for measuring genuine general relativistic effects come from gravitational waves, a direct measure of which would be the final sought for validation of theory. However, for this, Earth-based interferometers like LIGO \cite{Abbott:2006zx,Aasi:2014jkh} and VIRGO  \cite{TheVirgo:2014hva}, are
raising more and more their sensitivity, and there is a moderate optimistic perspective to observe gravitational wave signals in the next forthcoming years, especially those associated with coalescence scenarios in binary systems, i.e., from \lq\lq strong gravitational field" phenomena.

In the \lq\lq weak gravitational field," instead, general relativity enters only as a small modification of the Newtonian gravity, i.e., it constitutes an approximation of the full theory. Commonly used approximation schemes are 1) the so-called Post-Netwonian (PN) approximation which incorporates corrections to the flat Minkowski spacetime from a series expansion in powers of
$1/c^2$ (also called a ``slow-motion approximation''), where $c$ denotes the speed of light in vacuum (see, e.g., \cite{Damour:1990pi,Damour:1991yw,Damour:1992qi,Damour:1993zn}; recently PN has been improved by the effective-one-body (EOB) formalism, when dealing with binary systems \cite{Buonanno:1998gg}); 2) the so-called Post-Minkowskian (PM) approximation which incorporates corrections to the flat Minkowski spacetime from a series expansion in powers of $G$ (also called a ``fast-motion approximation''), where $G$ denotes the gravitational constant.

Among the  advantages to operate in the far- or weak-field regime is that one can still consider the gravitational theory as a \lq\lq linear" analogous of electromagnetism
and, as a consequence, discuss most of the gravitational phenomena according to their electromagnetic counterpart.
This very useful formalism, termed \lq\lq gravitoelectromagnetism," (GEM) \cite{Braginsky:1976rb,mas93} is not a at all a new theory, but only a convenient language to re-express gravitation
in a form which may be of some utility for what concerns our intuition. For example, introducing properly gravitoelectric and gravitomagnetic fields, one can discuss test particle motion (i.e., geodesic motion) as being (locally) a motion in presence of an external force, the latter formally represented by the  GEM analogue of the Lorentz force.

Without entering the details of the general GEM approach (which can even applied in the full exact theory~\cite{Jantzen:1992rg} and not only in its linearizations, or even at the Newtonian level), we limit our considerations here to the case of \lq\lq linear gravitoelectromagnetism" \cite{mas93}, which involves a PM approximation scheme (namely, it works at linear order in $G$).
In this context, we explore here
the role of corrections to the Lorentz gravitoelectromagnetic force due to a time-varying gravitomagnetic vector potential.
As we will show below, such corrections are represented by an acceleration contribution constituting the analogous of the induction law in electromagnetism, which we term as \lq\lq gravitational induction" acceleration term. Here, more specifically,  we will  evaluate the contribution of the  single nontrivial  gravitational induction  term on the mean orbital motion of a test particle, in the case of a gravitational field generated by a localized and slowly rotating astronomical source.

\section{Gravitational induction acceleration terms}

Let us consider the spacetime generated by a localized and slowly rotating astronomical source, referred to Cartesian-like coordinate system $x^\mu=(ct,{\mathbf r})$ with ${\mathbf r}=(x,y,z)$ and $\mu=0,1,2,3$. In the linear approximation, the spacetime metric generated by such a source can be written as $g_{\mu\nu}=\eta_{\mu\nu}+h_{\mu\nu}$, where $\eta_{\mu\nu}$ is the Minkowski metric (with signature $+2$, according to the convention used here)  and $h_{\mu\nu}$ is a first-order perturbation, often re-expressed by its  trace-reverse counterpart $\bar h_{\mu\nu}=h_{\mu\nu}-\frac12 h \eta_{\mu\nu}$ with $h={\rm tr}(h_{\mu\nu})$.
It is well known that imposing the Lorenz (or \lq\lq transverse") gauge condition
$\bar h^{\mu\nu}{}_{,\nu}=0$, the gravitational field equations can cast in the form
\beq
\label{eq:1}
\dal \bar h_{\mu\nu}=-\frac{16\pi G}{c^4}T_{\mu\nu}\ .
\eeq
The general solution of (\ref{eq:1}) is given by the   retarded solution
\beq
\label{eq:2}
 {\bar h}_{\mu\nu}=\frac{4G}{c^4}\int \frac{T_{\mu\nu}(ct-|{\mathbf r}-{\mathbf r}'|, {\mathbf r}')}{|{\mathbf r}-{\mathbf r}'|}\rmd^3 x'\ ,
\eeq
to which one can add a general solution of the homogeneous wave equation $\dal \bar h_{\mu\nu}=0$.
Neglecting all terms of $O(c^{-4})$, the metric tensor has components given by
\beq
\bar h_{00}=4\Phi/c^2\,, \quad
\bar h_{0i}=-2A_i/c^2\,,\quad \bar h_{ij}=O(c^{-4})\,,
\eeq
 where $\Phi(t,{\mathbf r})$ is termed  {\it gravitoelelctric} potential and ${\mathbf A}(t,{\mathbf r})$ is the {\it gravitomagnetic} vector potential \cite{mas93,Bini:2008cy} . Both potentials can be used to discuss gravity as linear analogy of electromagnetism, namely {\it gravitoelectromagnetism} (GEM). The spacetime metric results then
\beq
\label{eq:3}
\rmd s^2= -c^2 \left(1-2\frac{\Phi}{c^2}\right)\rmd t^2 -\frac4c ({\mathbf A}\cdot \rmd {\mathbf r})\rmd t +
 \left(1+2\frac{\Phi}{c^2}\right)\delta_{ij}\rmd x^i \rmd x^j\, .
\eeq
Denoting by $M$ and ${\mathbf S}$ the   mass and the angular momentum of the source, respectively,
the GEM potentials (in the region far from the source, characterized by $r\gg GM/c^2$ and $r\gg J/(Mc)$, with  $r=|{\mathbf r}|$) can be expressed as
\beq
\label{eq:4}
\Phi = \frac{GM}{r}, \qquad {\mathbf A}=\frac{G}{c}\frac{{\mathbf S}\times {\mathbf r}}{r^3}\ .
\eeq
Note that the above choice of $\Phi$ is fully consistent with the PM approximation of GR; differently, working in the PN approximation this choice
would not be fully consistent even at 1PN and one should  include  additional terms \cite{pois-will}, i.e.
\beq
\Phi = \frac{GM}{r} \left(1+\frac{GM}{c^2r}  \right)+O\left( \frac{1}{c^4} \right)\,.
\eeq
However, as stated above, $O(G^2)$ terms are not considered here.

The gauge condition $\bar h^{\mu\nu}{}_{,\nu}=0$ implies that
\beq
\label{eq:5}
\frac{1}{c}\partial_t \Phi+\nabla \cdot \left( \frac12 {\mathbf A}\right)=0\,,
\eeq
i.e., it is related to the conservation of energy-momentum of the source via Eq. (\ref{eq:1}). Indeed, let us denote
$T^{00}=\rho c^2$ and $T^{0i}=cj^i$, where $j^\mu=(c\rho,{\mathbf j})$ is the mass-energy current of the source; then, Eq. (\ref{eq:5}) can be cast in the form $j^\mu{}_{,\mu}=0$.
Similarly to the potentials, it is possible to define the gravitoelectric field ${\mathbf E}$ and the gravitomagnetic field ${\mathbf B}$
in analogy with electromagnetism
\beq
\label{eq:6}
{\mathbf E}=-\nabla \Phi-\frac{1}{c}\partial_t \left( \frac12 {\mathbf A}\right), \qquad {\mathbf B}=\nabla \times {\mathbf A}\,.
\eeq
It follows from these definitions that
\beq
\label{eq:7}
\nabla \times {\mathbf E}=-\frac{1}{c}\partial_t \left( \frac12 {\mathbf B}\right),\qquad
\nabla \cdot \left(\frac12 {\mathbf B} \right)=0\,,
\eeq
while the gravitational field equations (\ref{eq:1}) imply
\beq
\label{eq:8}
\nabla \cdot {\mathbf E}=4\pi G \rho, \qquad \nabla \times \left( \frac12 {\mathbf B}\right)=
\frac{1}{c}\partial_t {\mathbf E}+\frac{4\pi G}{c}{\mathbf j}.
\eeq
Eqs. (\ref{eq:7}) and (\ref{eq:8})  are the Maxwell-like equations for the GEM field and one can easily translate in the GEM framework most of the results of classical electrodynamics (see e.g., Ref. \cite{mas93} for a detailed discussion), by assuming the convention that the source has gravitoelectric charge $Q_E=GM$ and gravitomagnetic charge $Q_B=2GM$, while
a test particle of  mass $m$ has gravitoelectric charge $q_E=-m$ and gravitomagnetic charge $q_B=-2m$.
The opposite signs of $(q_E,q_B)$ with respect to those of $(Q_E,Q_B)$ are explained by the attractive nature of gravity; furthermore, the ratio of gravitomagnetic charge to the gravitoelectric charge is always 2, since the linear approximation of general relativity is characterized by a spin-2 field.

Let us consider the following simple choice of the GEM potentials (solution of the field equations)
\beq
\label{eq:23}
\Phi=\frac{GM}{r}, \qquad {\mathbf A}=\frac{G}{c}\frac{{\mathbf S}(t)\times {\mathbf r}}{r^3}\, .
\eeq
The geodesic motion of test particles in this spacetime is described by the equation \cite{Bini:2008cy}
\begin{eqnarray}
\label{eq:37new}
\frac{\rmd {\mathbf v}}{\rmd t}+\frac{GM{\mathbf r}}{r^3}&=& -\frac{2}{c}{\mathbf v}\times {\mathbf B}+\frac{2G}{c^2}\frac{\dot {\mathbf S}\times {\mathbf r}}{r^3}\,.
\end{eqnarray}

The purpose of the present analysis is to examine in detail the acceleration term involving $\dot{\mathbf S}$, namely
\beq
\label{gi_terms}
{\mathbf W}_{\dot S}^{(c^{-2})}=\frac{2G}{c^2}\frac{\dot {\mathbf S}\times {\mathbf r}}{r^3}\,.
\eeq
It is  associated with time-varying gravitomagnetic fields, and hence is responsible for eventual
gravitational induction effects on the orbital motion, at least in the linear approximation of the gravitational field considered here.
Furthermore, 
in the approximation considered here, agrees with the virial theorem constraints properly written in this case, as discussed in detail in the appendix  B.

\section{Calculation of the orbital effects}\label{orbef}

In order to obtain the cumulative orbital effects induced by a generic perturbing acceleration ${\mathbf W}$ whose analytical expression is known, the latter is customarily decomposed into three mutually orthogonal components; then, to first order in the perturbation, these components are evaluated along the unperturbed Keplerian ellipse, assumed as reference orbit, and inserted in the right-hand-sides of the Eq. (\ref{eq:37new}) to determine the variation of the associated orbital elements. Finally, the average over one full orbital revolution is taken.
We point out that  this procedure is general enough and it can be successfully applied to any disturbing acceleration, irrespectively of its physical origin.

\begin{figure}
\includegraphics[scale=0.5]{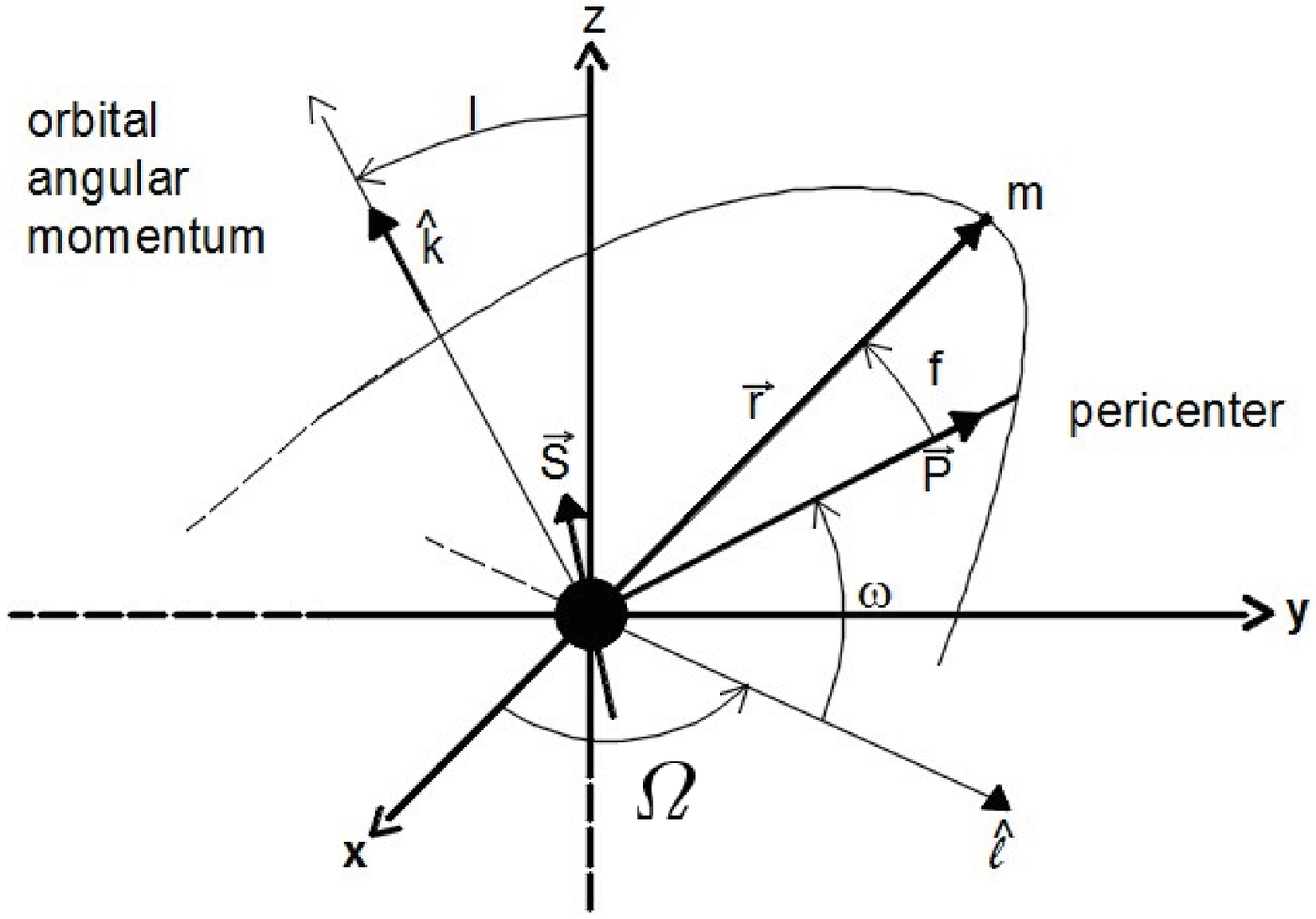}
\caption{\label{fig:1} Schematic representation of the geometrical elements of the orbit as well as of the notation used for main reference angles.}
\end{figure}

Following a standard notation \cite{brumberg} (shortly recalled below, for convenience; see also Ref. \cite{Iorio:2014zaa}, as well as the Tables I and II in Appendix A),  let us write the position and velocity vectors of the test particle in the form
\beq
\label{vecr}
{\mathbf r}  = r\left({\hat {\mathbf P} }\cos f + \hat{\mathbf Q}\sin f\right)\,, \quad r = \frac{p}{1+e\cos f},
\eeq
and
\beq
\label{velo}
{\mathbf v}  =  \sqrt{\frac{\mu_{\textcolor{black}{{\rm b}}}}{p}}\left[ -\hat {\mathbf P} \sin f + \hat{\mathbf Q}\ton{\cos f + e}\right].
\eeq
In (\ref{vecr}) and (\ref{velo}), $p=a(1-e^2)$ is the semilatus rectus of the particle; $e$ denotes the eccentricity; the unit vectors $\hat {\mathbf P}$ (directed along the line of the apsides, towards the pericenter of the particle) and $\hat{\mathbf Q}$ (directed transversally to the line of the apsides, in the orbital plane of the particle) read \cite{brumberg}
\begin{eqnarray}
\hat {\mathbf P}  & = \hat{\mathbf l}\co + \hat{\mathbf m}\so,  \qquad
\hat{\mathbf Q}  = -\hat{\mathbf l}\so + \hat{\mathbf m}\co,
\end{eqnarray}
where two other unit vectors $\hat {\mathbf l}$ (directed along the line of the nodes, towards the ascending node of the particle) and $\hat{\mathbf m}$ (directed transversally to the line of the nodes, in the orbital plane of the particle), conveniently used in the standard description of the motion.
Finally, ${\mathit \Omega}$ is the longitude of the ascending node of the particle, $\omega$ is the argument of the pericenter of the particle and $f$ is its true anomaly.
Fig. (\ref{fig:1})
represents (schematically) the geometrical elements of the orbit as used here.

Denoting by ${\mathit I}$ the inclination of the orbital plane of the particle, we can express
the Cartesian components of $\hat{\mathbf l}$ and $\hat{\mathbf m}$   in polar form as
\beq
\begin{array}{lllllllll}
{\hat{l}}_x &=&\cO\,,  &
{\hat{l}}_y &=&\sO\,,   &
{\hat{l}}_z &=&0\,,  \cr
{\hat{m}}_x  &=& -\cI\sO\,, &
{\hat{m}}_y  &=& \cI\cO\,,  &
{\hat{m}}_z  &=& \sI\,.
\end{array}
\eeq
Finally, the radial, transverse and normal components of ${\mathbf W}$ can, thus, be calculated as \cite{brumberg}
\begin{eqnarray}
\label{a_rtn}
W_R  ={\mathbf W} \cdot \hat{\mathbf r}\,, \quad
W_T  = {\mathbf W} \cdot \ton{\hat {\mathbf k} \times\hat{\mathbf r} }\,, \quad
W_N  = {\mathbf W} \cdot \hat {\mathbf k}\,,
\end{eqnarray}
where the Cartesian components of $\hat {\mathbf k}=\frac{{\mathbf r}\times {\mathbf v}}{|{\mathbf r}\times {\mathbf v}|}$ (unit vector of the orbital angular momentum per unit mass of the particle, ${\mathbf k}={\mathbf r}\times {\mathbf v}$) are given by
\begin{eqnarray}
{\hat{k}}_x =\sI\sO\,, \quad
{\hat{k}}_y  = -\sI\cO\,, \quad
{\hat{k}}_z  = \cI\,.
\end{eqnarray}

Eqs. (\ref{a_rtn}), evaluated along the unperturbed Keplerian ellipse (i.e., the unperturbed orbit, see Eq. (\ref{vecr})$_2$), should then be inserted into the right-hand-sides of Eqs. (\ref{eq:37new}). To the first order in the disturbing acceleration $W$, they are given by
\begin{eqnarray} \label{dadf}
\dert a t & =& \frac{2a^2}{\sqrt{\mu_{\textcolor{black}{{\rm b}}} p}}\qua{e \sin f W_R + \ton{\frac{p}{r}}W_T},   \nonumber \\
\dert e t & =& \sqrt{\frac{p}{\mu_{\textcolor{black}{{\rm b}}}}}\qua{\sin f W_R + \ton{1+\frac{r}{p}}\cos f W_T  +e\ton{\frac{r}{p}}W_T}, \nonumber \\
\dert I t &=& \frac{r \cos u W_N}{\sqrt{\mu_{\textcolor{black}{{\rm b}}} p} },  \nonumber \\
\dert {\mathit{\Omega}} t &=& \frac{r \sin u W_N}{\sI\sqrt{\mu_{\textcolor{black}{{\rm b}}} p}},  \nonumber \\
\dert \omega t & =& \frac{1}{e}\sqrt{\frac{p}{\mu_{\textcolor{black}{{\rm b}}}}}\qua{-\cos f W_R +\ton{1+\frac{r}{p} }\sin f W_T}-\cI\dert {\mathit{\Omega}}t\,,
\end{eqnarray}
where $u=\omega+f $ represents the argument of the particle's latitude.
In the same approximation, the long-term orbital rates of change are obtained by averaging the right-hand-sides of Eqs. (\ref{dadf}) over one orbital period $P_{\rm b}$ of the test particle by means of
\beq
\label{dtdf}
dt = \frac{r^2}{\sqrt{\mu_{\textcolor{black}{{\rm b}}} p}}df.
\eeq
Note that in Eq. (\ref{dtdf}), the instantaneous changes of $\mathit{\Omega}$ and $\omega$, induced by the perturbation itself, are neglected, consistently with the approximations considered here; see, e.g.,  \cite{Iorio:2014zaa} for details, as well as for a discussion about second-order and mixed effects due to the presence of more than one perturbing acceleration.

\subsection{The gravitational induction acceleration of order $O\ton{c^{-2}}$}
\label{giprima}

Let us study the gravitational induction acceleration term ${\mathbf W}_{\dot S}^{(c^{-2})}
$, defined in Eq. (\ref{gi_terms}).
It is useful to introduce the following notation
\begin{eqnarray}
{\mathcal{A}} &=& {\dot{S_z}}\cI +\sI\ton{{\dot{S_x}}\sO - {\dot{S_y}}\cO}\nonumber\\
{\mathcal{B}} &=& {\dot{S_z}}\sI +\cI\ton{{\dot{S_y}}\cO - {\dot{S_x}}\sO }\nonumber\\
{\mathcal{C}} &=& {\dot{S_x}}\cO + {\dot{S_y}}\sO\,.
\end{eqnarray}
The radial, transverse, normal components of Eqs. (\ref{gi_terms}) are
\begin{eqnarray}
W^{(c^{-2})}_R \label{AR} & =& 0,  \nonumber \\
W^{(c^{-2})}_T \label{AR} & =& \frac{2G\ton{1+e\cos f}^2 {\mathcal{A}} }{c^2 a^2\ton{1-e^2}^2},  \nonumber \\
W^{(c^{-2})}_N \label{AR} \nonumber & =& -\frac{2G\ton{1+e\cos f}^2}{c^2 a^2\ton{1-e^2}^2}\qua{ {\mathcal{B} }\cu - {\mathcal{C}} \su }.
\end{eqnarray}
The long-term rates of change of the osculating Keplerian orbital elements, defined as
\beq
\left\langle \dert X t  \right\rangle =\frac{\nk}{2\pi}\int_0^{2\pi} \frac{dX}{df} df\,,
\eeq
with $\nk = \sqrt{\mu_{\textcolor{black}{{\rm b}}} a^{-3}}$ the Keplerian mean motion of the  particle,
are conveniently expressed in terms of the orbital eccentricity $e$ (or better through the combination ${\mathcal E}=\sqrt{1-e^2}$), and are (straightforwardly) given by
\begin{eqnarray}
\label{dadt}
\left\langle \dert a t \right\rangle & = & \frac{4G {\mathcal{A} }}{c^2 a^2 {\mathcal E}^2\nk} \,, \\
\label{dedt}
\left\langle \dert e t \right\rangle & =& \frac{2G\ton{1-{\mathcal E} }  {\mathcal{A}} }{c^2 a^3 e\nk} \,, \nonumber\\
\label{dIdt}
\left\langle \dert I t \right\rangle  & =&\frac{G\left\{
-[(1-{\mathcal E})\coo +(1+{\mathcal E})] {\mathcal{B}}+(1-{\mathcal E})  {\mathcal{C} }\soo
\right\}}{c^2 a^3 {\mathcal E}(1+{\mathcal E}) \nk}\,,\nonumber\\
\label{dOdt}
\left\langle \dert {\mathit{\Omega}} t  \right\rangle   & =& \frac{G\csc I\{
-(1-{\mathcal E})   {\mathcal{B}  }\soo + [(1+{\mathcal E})-(1-{\mathcal E})\coo] {\mathcal{C} }
\}}{ c^2 a^3 e^2{\mathcal E}(1+{\mathcal E}) \nk}\,,
\nonumber\\
 \label{dodt}
 \left\langle   \dert \omega t  \right\rangle   & =&  \frac{G\cot I\left\{
(1-{\mathcal E})  {\mathcal{B}}  \soo - [(1+{\mathcal E})-(1-{\mathcal E})\coo]  {\mathcal{C}}
\right\}}{2c^2 a^3 e^2{\mathcal E}(1+{\mathcal E}) \nk}\,.
\nonumber
\end{eqnarray}

 In Eqs. (\ref{dadt}),  no a priori simplifying assumptions on both $e$ and $I$ were made; moreover, also the spatial orientation of ${\mathbf S}$ was  left arbitrary in the sense that the primary's angular momentum was not aligned to any particular direction. However, we assumed that $\dot{\mathbf S}$ stays constant over one orbital revolution of the test particle; as we will see in Section \ref{obs}, such an assumption is amply verified in all the astronomical and astrophysical systems of potential interest to put to the test the predictions of  Eqs. (\ref{dadt}).

Expanding in powers of $e$ up to $O(e^2)$, Eqs. (\ref{dadt}) reduce to
\begin{eqnarray}
\label{dadtapp}
\left\langle \dert a t  \right\rangle     &=& \frac{4G  {\mathcal{A}} }{c^2 a^2\nk}\,,\qquad \left\langle \dert e t \right\rangle     = \frac{Ge  {\mathcal{A} }}{c^2 a^3\nk}\,,\qquad \left\langle \dert I t \right\rangle   =\frac{G  {\mathcal{B}} }{c^2 a^3\nk}\,, \nonumber\\
\left\langle \dert {\mathit{\Omega}} t  \right\rangle  &=& \frac{G  {\mathcal{C}}\csc I}{c^2 a^3\nk}\,,\qquad
\left\langle \dert \omega t \right\rangle   = -\frac{G  {\mathcal{C} }\cot I }{c^2 a^3\nk}\,.
\end{eqnarray}
Eqs. (\ref{dadt}) and (\ref{dadtapp}) imply a non-vanishing long-term variation of the semimajor axis proportional to $a^{-1/2}$, while the other rates of change fall as $a^{-3/2}$. For circular orbits, obviously the eccentricity remains constant, as shown in Eq. (\ref{dadtapp}).

To some extent, eqs. (\ref{dadt}) and (\ref{dadtapp}) extend and generalize the results obtained in the literature. Indeed, in
\cite{Bini:2008cy,RuIo2010}, only the length $S$ of the primary's angular momentum ${\mathbf S}$, aligned constantly along the reference $z$ axis, was assumed time-dependent according to a particular law.  Moreover, the low-eccentricity approximation was used for the orbital configuration of the test particle. In  \cite{Bini:2008cy}, a different perturbative scheme and orbital parametrization were adopted, so that an explicit comparison with the present results is not straightforward.

As a final remark, we note that  the acceleration term $\frac{2}{c}{\mathbf v}\times {\mathbf B}$ in Eq. (\ref{eq:37new}) will not be further discussed here. Indeed,  assuming   ${\mathbf S}(t)$ slowly varying along the orbit (which is the approximation considered here), the standard Lense-Thirring precessions~\cite{Iorio:2010ah} are inferred. They exhibit a slow modulation due to the time dependence of ${\mathbf S}(t)$ characterized by timescales much longer than the test particle's orbital period.

\section{Confrontation with observations}\label{obs}

In general, temporal changes of the angular momentum ${\mathbf S}$ of an astronomical body can be due to variations of either its magnitude  $S$  and its orientation $\hat{\mathbf S}$. Thus, passing to a polar representation of ${\mathbf S}$, namely
\begin{eqnarray}
S_x \label{Sx} &= S\cos\alpha\cos\delta,  \quad
S_y \label{Sy}  = S\sin\alpha\cos\delta,  \quad
S_z \label{Sz} = S\sin\delta,
\end{eqnarray}
the rates of change of its components  can be written, in general, as
\begin{eqnarray}
\label{dSxdt}
{\dot{S_x}}  & =& \dot S\cos\alpha\cos\delta - S\ton{\dot\alpha\sin\alpha\cos\delta +\dot\delta\cos\alpha\sin\delta },  \nonumber \\
{\dot{S_y}}& =& \dot S\sin\alpha\cos\delta + S\ton{\dot\alpha\cos\alpha\cos\delta -\dot\delta\sin\alpha\sin\delta },  \nonumber \\
{\dot{S_z}}  & =& \dot S\sin\delta +S\dot\delta\cos\delta.
\end{eqnarray}
 In Eqs. (\ref{Sx}) and (\ref{dSxdt}),
$\alpha$ and $\delta$ are two angles characterizing the direction of $\hat{\mathbf S}$ in space: $\alpha$ is the longitude and $\delta$ is the latitude in some coordinate system. For example, if Celestial coordinates are adopted, as customarily in many practical observations and data reductions, $\alpha$ is the right ascension (RA) and $\delta$ is the declination (DEC). From Eqs. (\ref{dSxdt}), it turns out that, even if ${\mathbf S}$ is aligned with, say, the $z$ axis adopted ($\delta = 90^{\circ}$), so that $\dot S_z=\dot S$, in general ${\dot{S_x}}$ and ${\dot{S_y}}$ do not vanish because of $\dot\delta$ depending on $\alpha$.

Let us, now, look at some astronomical and astrophysical scenarios accessible to accurate observations in which temporal variations of ${\mathbf S}$ are known to occur.

\subsection{The Sun}\label{sole}
The spin angular momentum of the Sun, determined from helioseismology, amounts to \cite{Pijpers:1998eb}
\beq
S_{\odot}= 1.90 \times 10^{41}~\textrm{kg~m}^2~\textrm{s}^{-1}\,;
\eeq
in Celestial coordinates, its orientation is characterized by \cite{Archinal}
\begin{eqnarray}
\alpha_{\odot} \label{aSun} &= 286.13^{\circ}, \quad
\delta_{\odot}  = 63.87^{\circ}\,.
\end{eqnarray}
The Sun loses angular momentum because of the torque exerted by its magnetic field.
According to Ref. \cite{Weber:1967kc}, its rate of change can be written as
\beq
\label{dSSundt}
\dot S_{\odot} = -\frac{S_{\odot}}{\tau}, \qquad \tau=7\times 10^9~\textrm{yr}.
\eeq
As a consequence, Eqs. (\ref{dadt}), calculated by using (\ref{aSun})-(\ref{dSSundt}) in Eqs. (\ref{Sx}),
yield minute orbital effects for either Mercury or a hypothetical dedicated probe orbiting the Sun at a few Solar radii: suffice it to say that the semimajor axis would experience a decrease as little as $\approx 10^{-11}$ m yr$^{-1}$, while the precessional rates of the angular orbital elements would be of the order of $\approx 10^{-14}-10^{-15}$ milliarcseconds per year (mas yr$^{-1}$).

The Carrington elements  $\i_{\rm C},\Omega_{\rm C}$
are usually adopted to fix the orientation of the Sun's spin axis with respect to the ecliptic; latest measurements performed with the instrumentation onboard the dedicated SOHO spacecraft yield \cite{beck}
\begin{eqnarray}
\i_{\rm C} \label{IC} & = 7.55^{\circ},  \quad
\Omega_{\rm C} = 73.5^{\circ}.
\end{eqnarray}
The standard planetary N-body torques induce secular precession of ${\hat{\mathbf S}}_{\odot}$ given by \cite{Giles}
\begin{eqnarray}
\dot {\i}_{\rm C} & = 0, \quad
\dot \Omega_{\rm C} \label{dOCdt} & =  0.013958^{\circ}~\textrm{yr}^{-1}.
\end{eqnarray}
The resulting orbital effects from (\ref{dadt})-(\ref{dodt}), referred to the ecliptic through Eqs. (\ref{IC}) and (\ref{dOCdt}) inserted in Eq. (\ref{Sx}), are of the order of $\approx 10^{-6}$ m yr$^{-1}$ and $\approx 10^{-9}$ mas yr$^{-1}$, respectively.

\subsection{Jupiter}\label{juno}
In the Solar system, the planet exhibiting the largest angular momentum is Jupiter
\cite{Soffel}, with \beq S_{\jupiter}\approx 6.9\times 10^{38}~\textrm{kg}~\textrm{m}^2~\textrm{s}^{-1}.\eeq Its spatial orientation ${\kap}_{\jupiter}$ is characterized by \cite{Archinal}
\begin{eqnarray}
\alpha_{\jupiter} \label{RA} & \approx 268.056595^{\circ}, \quad
\delta_{\jupiter} \label{DEC}   \approx 64.495303^{\circ}.
\end{eqnarray}
 As far as the rate of change of its direction in space is concerned, it is \cite{Archinal}
\begin{eqnarray}
\dot\alpha_{\jupiter} \label{dRAdt}  & \approx -0.006499^{\circ}~\textrm{cy}^{-1}, \qquad
\dot\delta_{\jupiter} \label{dDECdt}  \approx 0.002413^{\circ}~\textrm{cy}^{-1}.
\end{eqnarray}
Thus, from Eqs. (\ref{RA})-(\ref{dDECdt},) applied to Eqs. (\ref{dSxdt}), it follows
that, for the forthcoming Juno mission to Jupiter \cite{Matousek}, Eqs. (\ref{dadt}) yield completely negligible effects. Indeed, the predicted rate of change of the semimajor axis is as little as $\approx 10^{-7}$ m yr$^{-1}$, while the other effects are of the order of $\approx 10^{-9}$ mas yr$^{-1}$.
No changes of its magnitude are currently known.

\subsection{The Earth}\label{terra}
The spin angular momentum of the Earth amounts to \cite{Matousek}
\beq S_{\oplus} = 5.86\times 10^{33}~\textrm{kg~m}^2~\textrm{s}^{-1}.\eeq
At the epoch J2000.0, its spatial orientation is characterized by \cite{Archinal}
\begin{eqnarray}
\alpha_{\oplus} \label{RAE}& = 0^{\circ}, \quad
\delta_{\oplus}  = 90^{\circ},   \quad
\dot\alpha_{\oplus}   =  -0.641^{\circ}~\textrm{cy}^{-1}, \quad
\dot\delta_{\oplus} \label{dDECdtE} = -0.557^{\circ}~\textrm{cy}^{-1}.
\end{eqnarray}
Thus, Eq. (\ref{Sx}), calculated with Eqs. (\ref{dadt}) and (\ref{RAE}), shows that the gravitoelectric inductive effects on a typical Earth's spacecraft in a generic orbital configuration are completely negligible, amounting to about $\approx 10^{-7}$ m yr$^{-1}$ and $\approx 10^{-8}$ mas yr$^{-1}$.
The consequences of the overall secular increase \cite{Petit}
\beq
\label{buba}
\frac{\dot {\mathcal{P}}_{\oplus}}{{\mathcal{P}}_{\oplus}} = 1.9\times 10^{-10}~\textrm{yr}^{-1}
\eeq
of the Earth's rotation period ${\mathcal{P}}_{\oplus}$ of tidal and non-tidal origin are even smaller. Indeed, from
\beq
\label{slowdownE}
\dot S_{\oplus}= -3.5\times 10^{16}~\textrm{kg~m}^2~\textrm{s}^{-2}\,,
\eeq
which can be straightforwardly be inferred from (\ref{buba}), it turns out that only the inclination and the node of an Earth's artificial satellite in an arbitrary orbit  experience non-vanishing secular precessions as little as $\ll 10^{-12}$ mas yr$^{-1}$.
\subsection{The Double Pulsar {\rm PSR J0737-3039A/B}}\label{doppia}
In the case of the double pulsar PSR J0737-3039A/B \cite{Burgay,Lyne}, the general relativistic geodetic precession \cite{Barker:1975ae,Boerner} comes into play \cite{Breton} in changing the orientation of the angular momentum of one of its components. Indeed, the spin angular momentum ${{\mathbf S}}_{\rm B}$ of B precesses at a rate
\beq
\label{Geodetic}
\Psi_{\rm B} = \ton{\frac{2\pi}{P_{\rm b}}}^{5/3}T_{\odot}^{2/3}\ton{\frac{1}{1-e^2}}\frac{{\mathfrak{m}}_{\rm B}\ton{4 {\mathfrak{m}}_{\rm A} + 3 {\mathfrak{m}}_{\rm B}}}{2\ton{{\mathfrak{m}}_{\rm A}+{\mathfrak{m}}_{\rm B}}^{4/3}}
\eeq
 around the total angular momentum of the system, which essentially coincides with the orbital angular momentum \cite{Breton}. In Eq. (\ref{Geodetic}),  ${\mathfrak{m}}_{\rm A},~{\mathfrak{m}}_{\rm B}$ are expressed in units of Solar masses. In view of the compactness of the system, such a  precessional rate is far larger than those occurring in the Solar system due to classical $\textrm{N}-$body torques; thus, in principle, it may play a role in the orbital dynamics through the gravitational induction investigated in this paper.  As such, we apply Eqs. (\ref{dadt}) to the double pulsar, aware of the fact that it is just an order-of-magnitude calculation to check if such a binary is worth of more refined analyses involving also two-body and self-gravitating effects which, at least in principle, may play a role in the gravitational inductive patterns. In \cite{Breton}, the reference $x$ axis is directed along the line-of-sight towards the Earth, while the reference $y-z$ plane coincides with the plane of the sky. The polar angles of ${{\mathbf S}}_{\rm B}$ to be inserted in Eqs. (\ref{Sx}) are, in this case, $\phi$, $\theta$ \cite{Breton}, where $\theta$ is a colatitude. To express the time evolution of the pulsar's spin axis, a coordinate system aligned with the orbital angular momentum was adopted \cite{Breton}. In it, the colatitude\footnote{In \cite{Breton}, the symbols $\delta$ and $\Omega_{\rm B}$ are used for the colatitude and the geodetic precession rate of ${{\mathbf S}}_{\rm B}$, respectively. To avoid confusion with the declination of the Celestial coordinates and the Carrington element of the Sun, here we have used the symbols $\xi$ and $\Psi_{\rm B}$.} $\xi$ and the longitude $\phi_{\rm so}$ of the spin axis with respect to the total angular momentum are used, with \cite{Breton}
\begin{eqnarray}
\xi &= \xi_0,  \qquad
\phi_{\rm so} = {\phi_{\rm so}}_0 -\Psi_{\rm B} t.
\end{eqnarray}
The relation among $\phi,\theta$ and $\xi,\phi_{\rm so}$ is given by \cite{Breton}
\begin{eqnarray}
\cos\theta & =& \cos\ton{90^{\circ}-I}\cos\xi-\sin\ton{90^{\circ}-I}\sin\xi\cos\phi_{\rm so},  \nonumber \\
\sin\phi & =& \frac{\sin\xi\sin\phi_{\rm so}}{\sin\theta}\,.
\end{eqnarray}
Since the system is almost edge-on, the $z$ axis and the total
angular momentum are almost perfectly coincident \cite{Breton}. Thus, it can be posed \cite{Breton}
\begin{eqnarray}
\theta &\approx \xi,\quad
\phi  \approx \phi_{\rm so}.
\end{eqnarray}
The measured values at the epoch May 2, 2006 are
\begin{eqnarray}
\theta_0 \label{pin} & = 130.02^{\circ}, \quad
\phi_0  = 51.21^{\circ}, \quad
\Psi_{\rm B} \label{pon}  = 4.77^{\circ}~\textrm{yr}^{-1}.
\end{eqnarray}
By inserting Eqs. (\ref{pin}) in Eqs. (\ref{Sx}), (\ref{dadt}) we find that the angular-type precessional rates amount to about $\approx 10^{-5}-10^{-6}$ mas yr$^{-1}$, while the amplitude of the rate of the semimajor axis is of the order of  $\approx 10^{-4}$ m yr$^{-1}$. It is just the case to recall that, in binary pulsar systems, the main mechanism yielding a steady orbital shrinking is due to gravitational radiation damping. The related rate of change of the semimajor axis is \cite{LL}
\beq
\label{damp}
\left\langle \dert a t \right\rangle = -\frac{64G^3 m_{\rm A}m_{\rm B}(m_{\rm A}+m_{\rm B})}{5c^5 a^3};
\eeq
in the case of the double pulsar,  (\ref{damp}) yields a reduction of its relative semimajor axis at a rate as large as $2.4$ m yr$^{-1}$, thus largely overwhelming the gravitomagnetic dynamo effect.

The orbital effects due to the temporal decrease of the magnitude of the spin angular momentum caused by the braking action of torques due to magnetic dipole radiation \cite{Gunn} are even smaller. Indeed, by posing
\beq
\dot S = -2\pi\mathcal{I}\frac{\dot {\mathcal{P}}}{{\mathcal{P}}^2}\,,
\eeq
where \cite{brumberg} $\mathcal{I}\approx 10^{38}$ kg m$^2$ is the pulsar's moment of inertia, assumed constant, while $P$ is the pulsar's spin period, it turns out \cite{Lyne}
\beq
\label{slowdown}
\dot S_{\rm B} = -7.8\times 10^{22}~\textrm{kg~m}^2~\textrm{s}^{-2}.
\eeq
Inserting (\ref{slowdown}) in Eqs. (\ref{dSxdt}), allows to infer from (\ref{dadt}) a decrease of the semimajor axis of A as little as
$10^{-11}$ m yr$^{-1}$, while the angular precessions are of the order of $\ll 10^{-10}$ mas yr$^{-1}$.
Here, we neglect the angular momentum loss due to gravitational radiation
since (\ref{gi_terms}) was obtained in a non-radiative scheme. However, it turned out to be  negligible.

\section{Discussion}

In this work we have analyzed  the motion of test particles in the metric of a localized and slowly rotating astronomical source.
Our description  is framed in the context of linear gravitoelectromagnetism, in the linear approximation of general relativity which involves PM corrections to the flat spacetime.
We have considered, in particular, the effect on the orbit of the particle of the so called  gravitational induction acceleration,
due to time-varying gravitomagnetic fields. Within a first-order perturbative approach, we analytically calculated the rates of change of the Keplerian orbital elements by averaging them over one orbital period of the test particle. The orientation of the primary's spin axis was assumed arbitrary, and neither the inclination nor the eccentricity of the particle's orbit were restricted to small values.

We have shown  that,  either for most of planetary sources (the Sun, Jupiter, the Earth),  or for binary systems (the binary Pulsar {\rm PSR J0737-3039A/B}), 
the resulting cumulative effects on their orbit are completely negligible.

\appendix

\section{Notations}
The basic notations and definitions of orbital mechanics used in the text are summarized below in Table 1 and Table 2.
\begin{table}[h]
  \caption{Parameters of the source (primary body)\label{tab:pnorder}}
  \begin{center}
      \begin{tabular}{l|l}
\hline
$M $& mass of the primary \cr
$\mu_{\textcolor{black}{{\rm b}}}=GM $ & gravitational parameter of the primary\cr
$T=\mu_{\textcolor{black}{{\rm b}}} c^{-3} $ & gravitational time constant of the primary\cr
$\mathcal{I} $ & moment of inertia of the primary\cr
$\mathcal{P} $ & rotational period of the primary\cr
$S $ & angular momentum of the primary\cr
$\hat {\mathbf S} $ & unit vector of the spin axis of the primary\cr
\hline
 \end{tabular}
\end{center}
\end{table}

\begin{table}[h]
  \caption{Orbital parameters of the  particle (secondary body)\label{tab2}. }
  \begin{center}
      \begin{tabular}{l|l}
\hline
$a$  & semimajor axis of the  particle \cr
$\nk = \sqrt{\mu_{\textcolor{black}{{\rm b}}} a^{-3}}$  &  Keplerian mean motion of the  particle \cr
$P_{\rm b} = 2\pi \nk^{-1}$ & Keplerian orbital period of the  particle \cr
$e$ & eccentricity of the  particle \cr
$p=a(1-e^2)$ & semilatus rectum of the  particle \cr
$I$ & inclination of the orbital plane of the  particle \cr
${\mathit{\Omega}}$  & longitude of the ascending node of the  particle \cr
$\omega$  & argument of pericenter of the  particle \cr
$f$ & true anomaly of the  particle \cr
$u=\omega + f$  & argument of latitude of the  particle \cr
\hline
\end{tabular}
\end{center}
\end{table}

\section{Virial theorem constraints}
In  Mechanics (classical, relativistic) the virial theorem provides an equation  relating the (properly defined) average over time of the total kinetic energy  of a stable system bound by potential forces, with that of the total potential energy. In the case of interest here, besides the standard gravitational potential force, there exist other, spin-dependent, forces yielding  modifications to the more familiar virial result valid in the central field of a non-rotating body.
To see this, let us consider Eq. (\ref{eq:37new}) formally rewritten as
\beq
\label{force_eq_new}
m \frac{\rmd }{\rmd t} {\mathbf v}={\mathbf F}\,,
\eeq
i.e., with
\begin{eqnarray}
{\mathbf F} &=& -\frac{GmM{\mathbf r}}{r^3}-\frac{2m}{c}{\mathbf v}\times {\mathbf B}
+\frac{2Gm}{c^2}\frac{\dot {\mathbf S}\times {\mathbf r}}{r^3}\,.
\end{eqnarray}

Scalar multiplication by ${\mathbf r}$ of both sides of Eq. (\ref{force_eq_new}) leads to
\beq
m \left[\frac{\rmd }{\rmd t}({\mathbf r} \cdot {\mathbf v})  - v^2\right]={\mathbf F} \cdot {\mathbf r}\,,
\eeq
that is
\beq
\label{power_eq}
m \frac{\rmd }{\rmd t}({\mathbf r} \cdot {\mathbf v})=m v^2 + {\mathbf F} \cdot {\mathbf r}\,.
\eeq
For any quantity $X(t)$ one defines next the time average along the motion as
\beq
\langle X \rangle_\infty =\lim _{{\mathcal T}\to \infty}\frac{1}{{\mathcal T}}\int_0^{\mathcal T} X(t) dt\,.
\eeq
The time average of the left hand side of Eq. (\ref{power_eq}) in this sense implies
\beq
\langle m \frac{\rmd }{\rmd t}({\mathbf r} \cdot {\mathbf v}) \rangle_\infty = m \lim _{{\mathcal T}\to \infty}\frac{1}{{\mathcal T}} \left[  ({\mathbf r} \cdot {\mathbf v})_T-({\mathbf r} \cdot {\mathbf v})_0 \right]=0\,,
\eeq
which is certainly true for the case under consideration here of confined motion to some finite region so that $({\mathbf r} \cdot {\mathbf v})$ is never infinite.
Therefore,
\beq
\label{new_virial}
m\langle  v^2 \rangle_\infty  + \langle {\mathbf F} \cdot {\mathbf r}\rangle_\infty =0
\eeq
Eq. (\ref{new_virial}) represents the virial theorem in our case, with
\begin{eqnarray}
{\mathbf F}\cdot {\mathbf r}  &=& -\frac{GmM}{r}
-\frac{2m}{c}{\mathbf k}\cdot {\mathbf B}\,,
\end{eqnarray}
where we have introduced  the angular momentum per unit mass ${\mathbf k}={\mathbf r}\times {\mathbf v}$ (see Eq. (\ref{a_rtn}) and associated discussion) and hence,
\begin{eqnarray}
\langle  v^2 - \frac{G M}{r} \rangle_\infty &=&
\frac{2}{c}\langle {\mathbf k}\cdot {\mathbf B}\rangle_\infty\,.
\end{eqnarray}
This result shows, for example, that $\langle  v^2 \rangle_\infty \not \simeq  \langle \frac{G M}{r} \rangle_\infty$ as one might expect intuitively on the basis of familiar classical results, but it contains additional terms involving the spin ${\mathbf S}$
all within the limits of validity of the description of the field of an insulated spinning body as discussed in the present paper.

\section*{Acknowledgments}
We  are indebted with Prof. B. Mashhoon for useful discussion.
DB thanks the ICRANet and the  INFN (section of Naples) for partial support.


\begin{thebibliography}{00}

\bibitem{Abbott:2006zx}
  B.~Abbott {\it et al.}  [LIGO Collaboration],
  ``Searching for a Stochastic Background of Gravitational Waves with LIGO,''
  Astrophys.\ J.\  {\bf 659}, 918 (2007)
  [astro-ph/0608606].

\bibitem{Aasi:2014jkh}
  J.~Aasi {\it et al.}  [LIGO Scientific and VIRGO Collaborations],
   ``Searching for stochastic gravitational waves using data from the two colocated LIGO Hanford detectors,''
  Phys.\ Rev.\ D {\bf 91}, no. 2, 022003 (2015)
  [arXiv:1410.6211 [gr-qc]].

\bibitem{TheVirgo:2014hva}
  F.~Acernese {\it et al.}  [VIRGO Collaboration],
  ``Advanced Virgo: a second-generation interferometric gravitational wave detector,''
  Class.\ Quant.\ Grav.\  {\bf 32}, no. 2, 024001 (2015)
  [arXiv:1408.3978 [gr-qc]].

 \bibitem{Damour:1990pi}
  T.~Damour, M.~Soffel and C.~m.~Xu,
   ``General relativistic celestial mechanics. 1. Method and definition of reference systems,''
  Phys.\ Rev.\ D {\bf 43}, 3272 (1991).

 \bibitem{Damour:1991yw}
  T.~Damour, M.~Soffel and C.~m.~Xu,
   ``General relativistic celestial mechanics. 2. Translational equations of motion,''
  Phys.\ Rev.\ D {\bf 45}, 1017 (1992).

 \bibitem{Damour:1992qi}
  T.~Damour, M.~Soffel and C.~m.~Xu,
   ``General relativistic celestial mechanics. 3. Rotational equations of motion,''
  Phys.\ Rev.\ D {\bf 47}, 3124 (1993).

 \bibitem{Damour:1993zn}
  T.~Damour, M.~Soffel and C.~m.~Xu,
   ``General relativistic celestial mechanics. 4: Theory of satellite motion,''
  Phys.\ Rev.\ D {\bf 49}, 618 (1994).

\bibitem{Buonanno:1998gg}
  A.~Buonanno and T.~Damour,
   ``Effective one-body approach to general relativistic two-body dynamics,''
  Phys.\ Rev.\ D {\bf 59}, 084006 (1999)
  [gr-qc/9811091].

\bibitem{Braginsky:1976rb}
  V.~B.~Braginsky, C.~M.~Caves and K.~S.~Thorne,
   ``Laboratory Experiments to Test Relativistic Gravity,''
  Phys.\ Rev.\ D {\bf 15}, 2047 (1977).

\bibitem{mas93}
B.~Mashhoon,
``On the gravitational analogue of Larmor's theorem,''
Phys. Lett. A   {\bf 173}, 347 (1993).

\bibitem{Jantzen:1992rg}
  R.~T.~Jantzen, P.~Carini and D.~Bini,
   ``The Many faces of gravitoelectromagnetism,''
  Annals Phys.\  {\bf 215}, 1 (1992)
  [gr-qc/0106043].

\bibitem{Bini:2008cy}
  D.~Bini, C.~Cherubini, C.~Chicone and B.~Mashhoon,
   ``Gravitational induction,''
  Class.\ Quant.\ Grav.\  {\bf 25}, 225014 (2008)
  [arXiv:0803.0390 [gr-qc]].

\bibitem{pois-will}
E. Poisson and C.M. Will
``Gravity. Newtonian, Post-Newtonian, Relativistic,"
Cambridge University Press, Cambridge, UK, (2014).

\bibitem{Iorio:2010ah}
  L.~Iorio,
  ``General relativistic spin-orbit and spin-spin effects on the motion of rotating particles in an external gravitational field,''
  Gen.\ Rel.\ Grav.\  {\bf 44}, 719 (2012)
  [arXiv:1012.5622 [gr-qc]].

\bibitem{brumberg}
V.~A.~Brumberg,
``Essential Relativistic Celestial Mechanics,''
Ed. Adam Hilger, Bristol, (1991).

\bibitem{Iorio:2014zaa}
  L.~Iorio,
   ``Post-Newtonian direct and mixed orbital effects due to the oblateness of the central body,''
  \textcolor{black}{Int.\ J.\ Mod.\ Phys.\ D\ {\bf 24}, 1550067  (2015)} [arXiv:1402.5947].

\bibitem{RuIo2010}
M.~L.~Ruggiero and L.~Iorio, `` Gravitomagnetic time-varying effects on the motion of a test particle, ''
 Gen.\ Rel.\ Grav.\  {\bf 42}, 2393 (2010)
  [arXiv:0906.1281].

\bibitem{Pijpers:1998eb}
  F.~P.~Pijpers,
   ``Helioseismic determination of the solar gravitational quadrupole moment,''
  Mon.\ Not.\ Roy.\ Astron.\ Soc.\  {\bf 297}, 76 (1998)
  [astro-ph/9804258].

\bibitem{Archinal}
B.~A.~Archinal, {\it et al.}
``Report of
the IAU Working Group on Cartographic Coordinates and Rotational Elements: 2009,''
Celest.\ Mech.\ Dyn.\ Astr. {\bf 109}, 101 (2011).

\bibitem{Weber:1967kc}
  E.~J.~Weber and L.~Davis, Jr.,
   ``The Angular Momentum of the Solar Wind,''
  Astrophys.\ J.\  {\bf 148}, 217 (1967).

\bibitem{beck}
J.~G.~Beck and P.~Giles,
``Helioseismic Determination of the Solar Rotation Axis,"
Astrophys.\ J.\ Lett. {\bf 621}, 153 (2005).

\bibitem{Giles}
P.~M.~Giles,
``Time-Distance Measurements of Large-Scale Flows in the
Solar Convection Zone,''
Ph.D. Dissertation, Stanford University, (1999).

\bibitem{Soffel}
M.~Soffel, {\it et al.},
``The IAU 2000 Resolutions for Astrometry, Celestial
Mechanics, and Metrology in the Relativistic Framework:
Explanatory Supplement,''
Astron. J. {\bf 126}, 2687 (2003)
 [astro-ph/0303376].

\bibitem{Matousek}
S.~Matousek,
``The Juno New Frontiers mission,''
Acta \ Astronaut. {\bf 61}, 932 (2007).

\bibitem{Petit}
G.~Petit, {\it et al.},  
``IERS Conventions (2010)," IERS
Technical Note {\bf 36}, 1 (2010).

\bibitem{Stephenson}
F.~R.~Stephenson and L.~V.~Morrison,
``Long-Term Fluctuations in the Earth's Rotation: 700 BC to AD 1990,''
Proc.\ Roy.\ Soc.\ Lond. A\ Mat. {\bf 351}, 165 (1995).

\bibitem{Burgay}
M. Burgay, {\it et al.},
``An increased estimate of the merger rate of double neutron stars from observations
of a highly relativistic system,''
Nature {\bf 426}, 531 (2003)
[astro-ph/0312071].

\bibitem{Lyne}
A. G. Lyne, {\it et al.},
``A Double-Pulsar System: A
Rare Laboratory for Relativistic Gravity and Plasma Physics,''
Science {\bf 303}, 1153 (2004)
[astro-ph/0401086]

\bibitem{Barker:1975ae}
  B.~M.~Barker and R.~F.~O'Connell,
  ``Gravitational Two-Body Problem with Arbitrary Masses, Spins, and Quadrupole Moments,''
  Phys.\ Rev.\ D {\bf 12}, 329 (1975).

\bibitem{Boerner}
G.~Boerner, J.~Ehlers, and E.~Rudolph,
``Relativistic spin precession in two-body systems,''
Astron. Astrophys. {\bf 44}, 417 (1975)

\bibitem{Breton}
R.~P.~Breton, {\it et al.},
``Relativistic Spin Precession in the Double Pulsar,''
Science {\bf 321}, 104 (2008)
[arXiv:0807.2644 [astro-ph]].

\bibitem{LL}
L.~D.~Landau and E.~M.~Lifshitz,
``The classical theory of Fields,''
Ed. Pergamon Press, Oxford (1951).

\bibitem{Gunn}
J.~E.~Gunn and J.~P.~Ostriker,
``On the Nature of Pulsars. III. Analysis of Observations,"
Astrophys. J. {\bf 160}, 979 (1970).

\bibitem{Lorimer}
D.~Lorimer and M.~Kramer,
``Handbook of Pulsar Astronomy,''
Ed. Cambridge University Press, Cambridge (2005).

\bibitem{Matt}
S. P. Matt, {\it et al.},  
``The Mass-Dependence of Angular Momentum Evolution in Sun-Like Stars,"
ApJ {\bf 799}, L23 (2015)
[arXiv:1412.4786 [astro-ph]].

\end{thebibliography}
\end{document}